\documentclass[12pt]{article}

\newif\ifxelatex
\ifx\XeTeXversion\undefined
  \xelatexfalse
\else
  \xelatextrue
\fi

\ifxelatex
  \usepackage{xeCJK}
\else
  \usepackage[utf8]{inputenc}
  \usepackage{CJKutf8}
  \AtBeginDocument{\begin{CJK}{UTF8}{bsmi}}  
  \AtEndDocument{\end{CJK}}
\fi

\usepackage{geometry}
\usepackage{csquotes}
\usepackage{amsmath, amssymb}
\usepackage{graphicx}
\usepackage{float}
\usepackage[font=small]{caption}
\usepackage{subcaption}
\usepackage{cite}
\usepackage{hyperref}

\usepackage[ruled,vlined]{algorithm2e}

\title{Tree-Based Text Retrieval via Hierarchical Clustering in RAG
 Frameworks: Application on Taiwanese Regulations}
\author{
Chia-Heng Yu \\
National Chengchi University \\
\texttt{arthur422tp@gmail.com} \\
\\ 
Yen-Lung Tsai \\
National Chengchi University \\
\texttt{yenlung@nccu.edu.tw}
}

\date{}

\begin{document}
\maketitle

\begin{abstract}
\noindent
Traditional Retrieval-Augmented Generation (RAG) systems employ brute-force inner product search to retrieve the top-k most similar documents, then combined with the user query and passed to a language model. This allows the model to access external knowledge and reduce hallucinations. However, selecting an appropriate k value remains a significant challenge in practical applications: a small k may fail to retrieve sufficient information, while a large k can introduce excessive and irrelevant content. To address this, we propose a hierarchical clustering-based retrieval method that eliminates the need to predefine k. Our approach maintains the accuracy and relevance of system responses while adaptively selecting semantically relevant content.\\
\noindent\\
In the experiment stage, we applied our method to a Taiwanese legal dataset with expert-graded queries.  
The results show that our approach achieves superior performance in expert evaluations and maintains high precision while eliminating the need to predefine k, demonstrating improved accuracy and interpretability in legal text retrieval tasks.\\
\noindent\\
Our framework is simple to implement and easily integrates with existing RAG pipelines, making it a practical solution for real-world applications under limited resources.
\end{abstract}

\section{Introduction}

We propose a hierarchical clustering-based retrieval method that eliminates the need to predefine k. By organizing text embeddings into a tree structure and applying a cosine-distance-based BFS locates the most relevant node. This design enables adaptive retrieval: abstract queries retrieve higher-level nodes, while specific ones reach individual leaf nodes. We evaluated on a Taiwanese legal corpus with expert-graded questions, our method outperforms traditional retrieval in both adjusted F1 score ($\beta=4$) and expert ratings, demonstrating significant improvements in precision and interpretability. The proposed framework is scalable and domain-adaptable, offering a practical alternative for knowledge-intensive applications under limited resources.\\
\\
We summarize our contributions as follows:
\begin{itemize}
    \item\textbf{No Need to Set k-value:}\\
We propose a Hierarchical Clustering tree-based retrieval method that removes the need to predefine the $k$ value, overcoming a common bottleneck in traditional RAG ~\cite{10.5555/3495724.3496517}  pipelines.\\
    \item\textbf{Linguistic Hierarchy:}\\
Our hierarchical clustering tree not only groups semantically similar documents, but also reflects the multi-level semantic structure of the corpus. Higher-level nodes represent broader concepts, while lower-level nodes retain concrete information. This enables flexible retrieval across various scenarios, offering interpretable and structured access to complex corpora.
\end{itemize}

We also make our code available in Github \footnote{\url{https://github.com/arthur422tp/hierachical}. Detailed instructions for installation and usage are provided in the repository.}.

\section{Related Work}

Recent advancements in Large Language Models (LLMs) based on the Transformer architecture ~\cite{vaswani_attention_2017} have significantly transformed the field of natural language processing (NLP).\\
\\  
While LLM have demonstrated strong performance across NLP tasks, they remain prone to hallucinations~\cite{10.1145/3703155}, often generating content that is inaccurate or fabricated. To address this limitation, Retrieval-Augmented Generation (RAG) augments LLMs with external corpora, grounding outputs in factual sources through retrieval modules\\
\\
RAG, whether based on Dense Passage Retrieval (DPR) ~\cite{karpukhin-etal-2020-dense} or inner product search, rely on a fixed $k$ value during retrieval, which can lead to suboptimal results due to choosing an inappropriate $k$. While some RAG pipelines adopt reranking mechanisms ~\cite{qin-etal-2024-large}~\cite{xiong2020multihop} or setting thresholds in a two-stage process to refine the retrieved results, relatively few efforts have focused on revising the retrieval phase itself to adaptively select relevant content based on semantic structure.\\
\\
In certain domain-specific RAG applications, such as legal or medical fields, selecting an appropriate $k$ value becomes particularly challenging due to the complexity and variability of the content. To address this issue, we propose a hierarchical clustering-based retrieval framework that dynamically determines the number of relevant documents by leveraging the semantic structure of both the query and the corpus.\\
\\
Related work includes RAPTOR~\cite{sarthi2024raptor}, which constructs a tree-based representation by clustering passages using Gaussian Mixture Models and performing recursive summarization with an LLM. Graph RAG~\cite{edge2024graphrag}, extracts entities and relations by LLM to build the knowledge graph, followed by hierarchical community detection using the Leiden algorithm. At query time, it applies a map-reduce strategy over community summaries to generate global answers.\\
\\
Both methods enhance the retrieval process by imposing structure onto the corpus, thereby enabling more targeted or informative responses from the LLM. Although these approaches are effective, they introduce significant computational overhead and added complexity to the retrieval pipeline. In particular, both RAPTOR and Graph RAG require extensive preprocessing to construct structured representations of the corpus—RAPTOR through recursive LLM summarization over clustered passages, and GraphRAG through LLM-based entity and relation extraction followed by graph construction and community detection, making them more costly and time-consuming during preprocessing.\\
\\
\textbf{In contrast}, our method builds a semantic tree directly from text chunks and their representative vectors, entirely avoiding LLM involvement in the structure-building process. This leads to a lightweight, easily deployable retrieval framework that preserves semantic structure while remaining efficient and scalable.

\section{Methodology}

    We employ hierarchical clustering with cosine distance as the clustering metric, organizing text vectors into a tree diagram through bottom-up. To enhance the interpretability of the retrieval process, we construct a hierarchical retrieval tree with index for our retrieval system.\\
    \\
    This approach ensures that the retrieval not only maintains high precision but also offers more comprehensive and structured results.

    \begin{figure}[htbp]
        \centering
        \includegraphics[width=5.5cm]{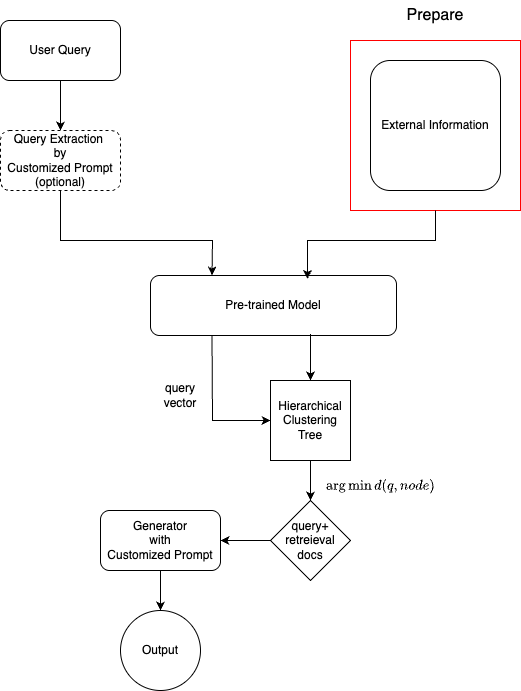}
        \caption{Flow of Hierarchical Clustering Tree}
        \label{fig:nn}
    \end{figure}
    
\subsection{Hierarchical clustering}
\subsubsection{Metric in Hierarchical clustering}
    We select cosine distance as our metric, given $\mathbf{v}, \mathbf{w}$ be any vector in a normed vector space $V$, the cosine distance is defined as follow:

    $$d(\mathbf{v},\mathbf{w}) =1-\frac{\langle\mathbf{v},\mathbf{w}\rangle}{|\mathbf{v}||\mathbf{w}|}$$ 
    where $|\cdot|$ is the Euclidean distance.\\
    \\
    Clearly, cosine distance is not a true distance, as it violates the triangle inequality. However, it preserves other properties of a distance metric and remains a valuable tool for text similarity.
    \begin{figure}[H]
        \centering
        \includegraphics[width=6.5cm]{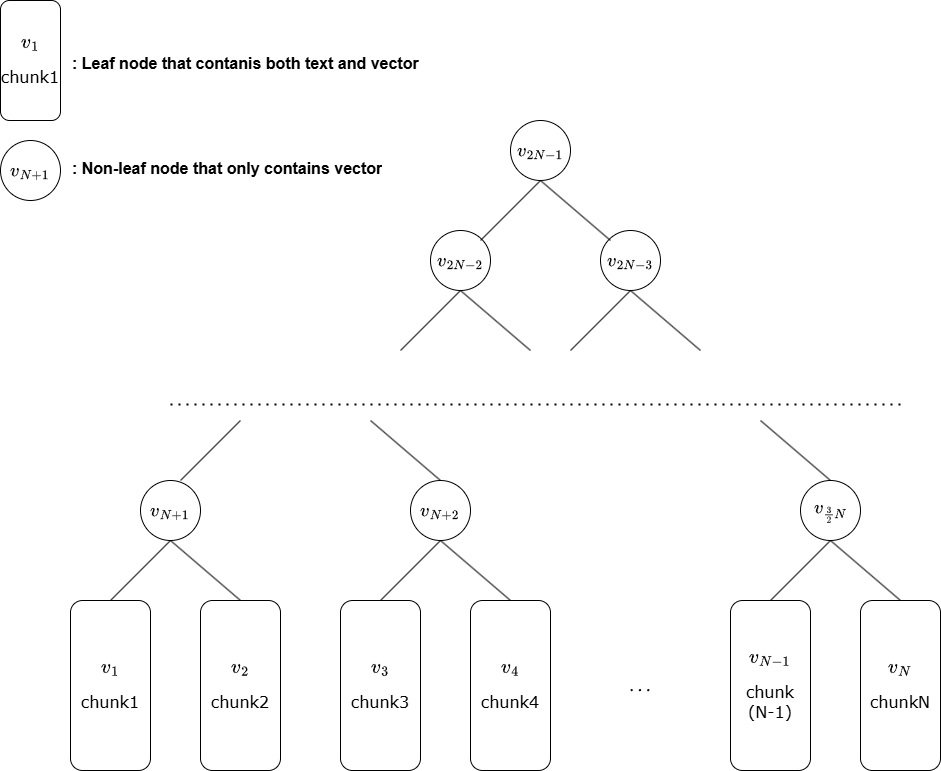}
        \caption{Hierarchical Clustering using Single-Linkage and Cosine Similarity. For each text vector, we initialize it as a separate cluster. We then iteratively merge the two closest clusters by cosine similarity until all vectors are merged into a single cluster, only the leaf nodes contains both vector and text; For the non-leaf nodes, we store only the vector, where the represent vector is determined by the mean of the vector of it children nodes.}
        \label{fig:nn}
    \end{figure}

\subsubsection{Construction}

As we select the metric for Hierarchical clustering, we are going to introduce how we construct the tree diagram.\\
\\
Given a set of \(N\) vectors \(\{\mathbf{v}_1, \mathbf{v}_2, \dots, \mathbf{v}_N\} \subseteq \mathbb{R}^d\), we aim to construct a hierarchical clustering tree using Single-Linkage with the cosine metric. The algorithm proceeds as follows:

\begin{algorithm}[H]
\caption{Hierarchical Clustering using Single-Linkage and Cosine Similarity}
\label{alg:hierarchical-clustering}
\KwIn{A set of text vectors $\{\mathbf{v}_1, \dots, \mathbf{v}_N\}$}
\KwOut{A hierarchical clustering tree}

Initialize each cluster $\mathcal{C}_i = \{\mathbf{v}_i\}$ with index $i$\;
Let $\mathcal{T} = \{\mathcal{C}_1, \dots, \mathcal{C}_N\}$ be the set of all clusters\;
Assign node index $\text{node}_i = i$ for each cluster\;

\While{$|\mathcal{T}| > 1$}{
    Compute cosine distance between all clusters:\newline
    \hspace*{1em}$d(\mathcal{C}_i, \mathcal{C}_j) = \min_{\mathbf{v}_a \in \mathcal{C}_i, \mathbf{v}_b \in \mathcal{C}_j} \left( 1 - \frac{\mathbf{v}_a \cdot \mathbf{v}_b}{\|\mathbf{v}_a\| \|\mathbf{v}_b\|} \right)$\;

    Identify clusters $\mathcal{C}_i$ and $\mathcal{C}_j$ with minimal distance\;
    Merge them: $\mathcal{C}_k = \mathcal{C}_i \cup \mathcal{C}_j$\;
    Assign index $\text{node}_k = N + m$, where $m$ is the number of merges so far\;
    Update $\mathcal{T}$ by replacing $\mathcal{C}_i, \mathcal{C}_j$ with $\mathcal{C}_k$\;
    Set cluster vector $\mathbf{v}_k = \frac{1}{|\mathcal{C}_k|} \sum_{\mathbf{v} \in \mathcal{C}_k} \mathbf{v}$\;
}
\end{algorithm}

\subsection*{Output}
The algorithm outputs a \textbf{dendrogram} representing the hierarchical clustering of the text vectors. Each node in the dendrogram corresponds to a cluster, indexed sequentially from \( 1 \) to \( 2N-1 \), where the first \(N\) nodes represent individual text vectors and the remaining \(N-1\) nodes represent merged clusters. The root node \(\text{node}_\text{root}\) corresponds to the final merged cluster containing all \(N\) vectors, and its index is:
\[
\text{node}_\text{root} = 2N - 1.
\]

\subsection{Searching Algorithm}

    We designed an algorithm to search our tree diagram, one of the algorithm is based on the index of all node in the retrieval tree, we search the most relevant node to the query. If such node is leaf node, then we output the text we stored in the leaf. if not, we collect its subtree and output all the text in the leaves of the subtree.\\
    \\
    To improve retrieval performance for overly abstract or vague queries, we introduce an optional query extraction module, where a language model reformulates the original query into a more concrete and structured representation. This is achieved using a CoT prompting ~\cite{10.5555/3600270.3602070} strategy, co-designed with domain experts, aimed at providing more consistent results.\\
    \\
    Additionally, in cases where the retrieved content remains overly broad, we optionally reapply Maximum Inner Product Search (MIPS) within the collected documents to further refine the results.\\
    \\
    This overall design ensures both interpretability and semantic consistency, leveraging the tree's hierarchical structure to dynamically adjust the retrieval granularity.\\

\begin{algorithm}[H]
\caption{BFS-Based Nearest Subtree Search Using Cosine Distance}
\label{alg:bfs-subtree}
\KwIn{Query vector $\mathbf{q}$, hierarchical tree with root node $root$}
\KwOut{Set of leaf text vectors $\mathcal{L}$ under the closest node}

Initialize a queue: $Q \gets \{root\}$\;
Initialize: $min\_distance \gets \infty$, $best\_node \gets root$\;

\While{$Q$ is not empty}{
    Dequeue the first node: $node \gets Q.pop()$\;
    
    Compute cosine distance between $\mathbf{q}$ and $node$’s vector representation $\mathbf{v}_{node}$:\newline
    \hspace*{1em}$d = 1 - \dfrac{\mathbf{q} \cdot \mathbf{v}_{node}}{\|\mathbf{q}\| \|\mathbf{v}_{node}\|}$\;

    \If{$d < min\_distance$}{
        $min\_distance \gets d$\;
        $best\_node \gets node$\;
    }

    \For{each child $c$ of $node$}{
        Enqueue $c$ into $Q$\;
    }
}
Collect all leaf text vectors $\mathcal{L}$ under $best\_node$\;
\Return $\mathcal{L}$\;
\end{algorithm}
        
\begin{figure}[htbp]
    \centering
    \includegraphics[width=7.5cm]{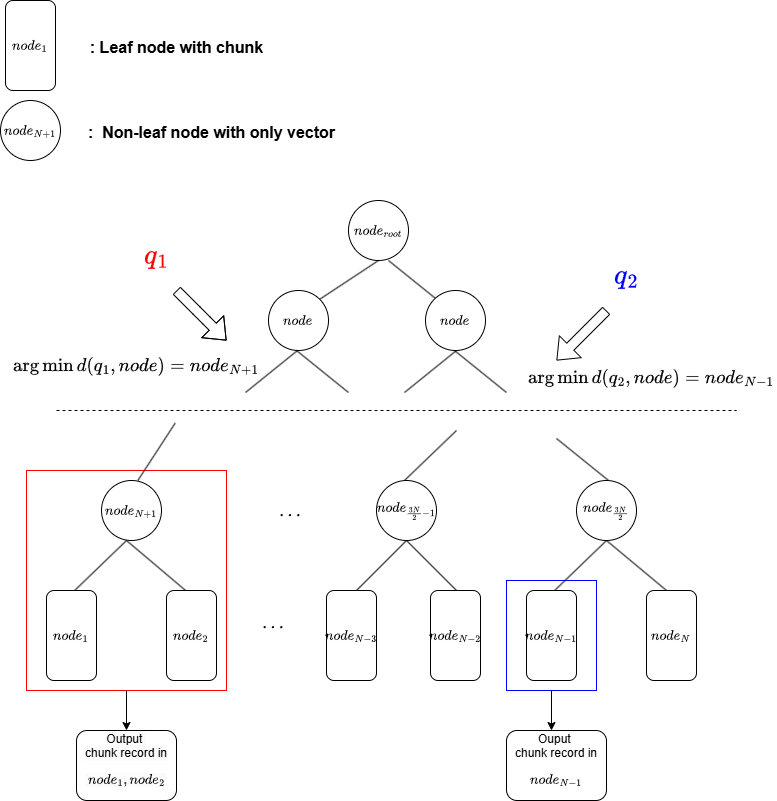}
    \caption{How to search in the retrieval tree. Given two queries  vector $q_1,q_2$, we can find the most relevant node in the retrieval tree. For $q_1$, the most relevant node is $node_{N+1}$, therefore let $node_{N+1}$ be the subroot of our retrieval tree, we output the text in the leaves of the subtree, which is $node_1$ and $node_2$; For $q_2$, the most relevant node is $node_{N-2}$, which is a leaf node, therefore we output the text in the leaf node directly.}
    \label{fig:nn}
\end{figure}

\paragraph{Time Complexity}
Both the construction and search algorithms are computationally efficient. The Hierarchical Clustering with Single-Linkage runs in $O(N^2)$ time due to pairwise distance computation, while the BFS-based search traverses at most $2N - 1$ nodes and $2N - 2$ edges in the worst case, resulting in $O(N)$ time complexity. These characteristics make our method lightweight and scalable for practical deployment..

\section{Experiment Setup}
In this section, we present how our we design our experiment to evaluate our retrieval system. We first introduce how we select and preprocess the corpus, the model selection, then how we evaluate different retrieval systems, and finally, we present the results of our evaluation.

\subsection{Corpus Selection and Preprocessing}
We select the Taiwan legal as our corpus for legal provision retrieval, we consulted legal experts and asked them to propose topics from chapters they specialize in, such as the General Principles of Civil Law and the Land Regulations.\\
\\
For the text preprocessing, we employed the text splitter module from the LangChain. For all experimental settings, we configured the splitter with a chunk size of 200 and a chunk overlap of 40, resulting in a 20$\%$ overlap. The parameter setting of overlap was motivated by heuristics recommended by senior engineers at Meta. For the setting of the chunk size, considered of the text density of law provisions, an excessively large chunk size may reduce retrieval accuracy since a single chunk may contain multiple concept of legal provisions while a chunk size that is too small can lead to inefficiency. In the context of Chinese legal text retrieval, we set the chunk size to 200 based on extensive project experience and iterative try and error. This value reflects based on our own observations and lessons learned from prior implementations by others.

\subsection{Model Selection}
For the pre-trained encoder model, we select intfloat/multilingual-e5-large ~\cite{wang2024e5}, an open source multilingual model developed by Microsoft and available on HuggingFace.\\
\\
For the generator, we select GPT-4o-mini as our testing LM. GPT-4o-mini is a variant of the GPT-4 model ~\cite{openai2025gpt4o_mini} , which is a multi-language model developed by OpenAI. GPT-4o-mini is a smaller version of GPT-4, with 128k context length. We choose GPT-4o-mini as our model because it performed well across multiple benchmark evaluations, achieving a score of 82$\%$ on the MMLU test according to Aritficial Analysis website.

\subsection{Evaluation Metrics}
For the evaluation of our retrieval system, we compare three different methods: Traditional inner product search , our Hierarchical clustering tree retrieval, and Hierarchical clustering tree retrieval with a query extraction. We evaluate those methods by Expert rating and adjusted F1 score, which is emphasized more on recall, design by the following formula:

$$\text{F1}_{\text{adjusted}} = \frac{(1+\beta^2)\,\text{Precision}\,\cdot \text{Recall}}{\beta^2\,\text{Precision} + \text{Recall}}$$

where $\beta = 4$ is the weight of precision. The adjusted F1 score is designed to emphasize more on recall, which is more important in law retrieval.

\paragraph{Expert rating}A subjective evaluation method, where we ask individuals who have passed the national bar examination and are licensed attorneys to assist us with the evaluation. The questions we provided are sourced from examinations administered by National Taiwan University, National Chengchi University, Taipei University(These universities are widely regarded as having the top three law departments in Taiwan.), and the national bar exam.\\
We provided the evaluation standard as follows:

\begin{table}[H]
    \centering
    \small
    \begin{tabular}{|c|p{2.5cm}|p{8.5cm}|}
    \hline
    \textbf{分數} & \textbf{說明} & \textbf{具體標準} \\
    \hline
    1 分 & 完全無關 & 檢索結果與問題無關，甚至產生誤導性內容，顯示對問題主題缺乏理解，無法提供有效資訊。 \\
    \hline
    2 分 & 部分相關但錯誤較多 & 結果與問題主題有關聯，但關鍵資訊錯誤或混亂，內容表述缺乏精確性與條理性。 \\
    \hline
    3 分 & 準確完整但尚未達專業標準 & 結果內容正確且具邏輯性，涵蓋問題所需之主要資訊與法條，表達清楚，僅在論述深度或結構完整性上略有不足。 \\
    \hline
    4 分 & 高度專業且清楚 & 回答內容完全符合問題需求，具高度專業性，涵蓋條文、條理明晰，適當完整連結法律依據、術語、制度背景與相關要素，展現良好專業知識。 \\
    \hline
    5 分 & 卓越專業、具實務應用深度 & 回答全面精準，展現整合能力與策略性，具體情境應用與策略建議能力，內容表現出明晰且具法律理解與實務判斷力。 \\
    \hline
    \end{tabular}
    \caption{Expert rating standard.}
    \end{table}

and we set the total score of our rating system by:
$$\text{Total} = 0.5 \cdot \text{Expert} + 0.5 \cdot 5 \cdot \text{F1}_{\text{adjusted}}$$

\section{Conclusion}

To evaluate the performance of our retrieval system, we compare the three methods, The \textbf{brute-force inner product search} was set as the \textbf{baseline (Origin method)}. The second method employed \textbf{hierarchical clustering tree retrieval (Tree method)}, and \textbf{the hierarchical clustering tree retrieval with query extraction(Tree+QE method)} as the third method. We evaluate the performance of the three methods by the \textbf{adjusted F1 score} and \textbf{expert rating}. The results are shown in Figure 4.\\
\begin{figure}[H]
    \centering
    \includegraphics[width=10cm]{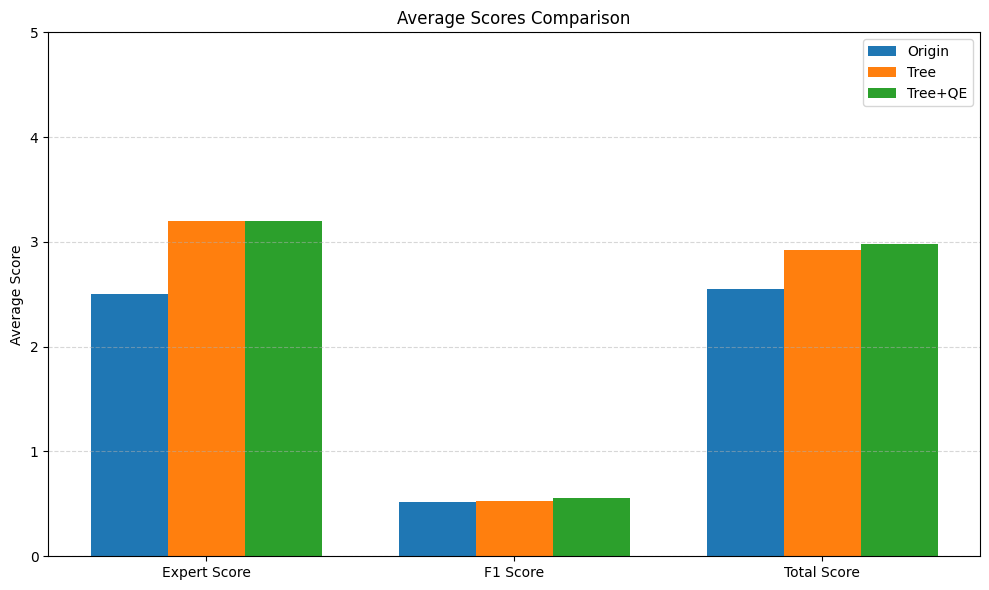}
    \caption{Average Expert Score, F1 Score, and Total Score for the three retrieval methods.}
    \label{fig:nn}
\end{figure}

In addition to the Average Score of the three methods, we also evaluate the distribution of score differences between each proposed method and the baseline.\\
\begin{figure}[htbp]
    \centering
    \includegraphics[width=10cm]{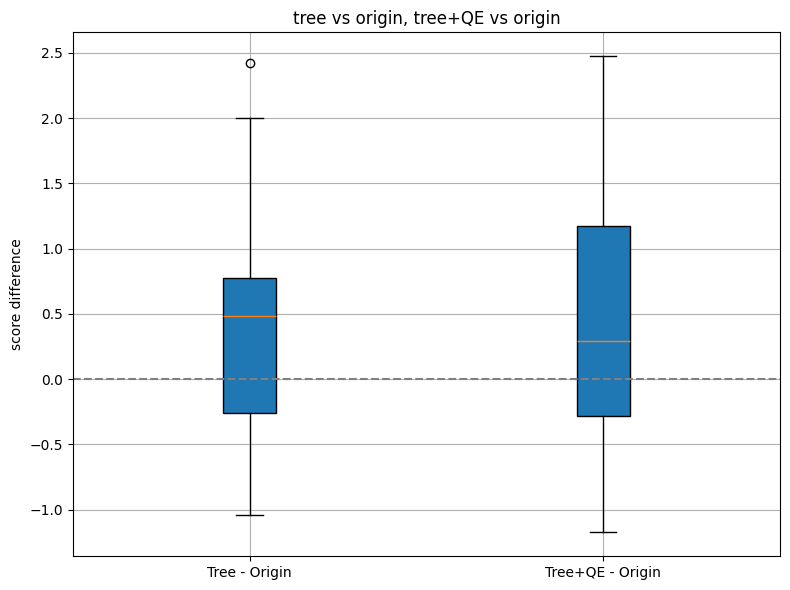}
    \caption{Box plot of the score differences between the Tree method and the Tree+Query method relative to the Origin method.}
    \label{fig:nn}
\end{figure}

As shown in Figure 5, the box plot presents the score differences of both \textbf{Tree method} and \textbf{Tree+Query methods} relative to the Origin method. While the Tree method demonstrates higher median improvement, its performance are more widely, with a greater spread and a outlier. In contrast, the Tree+Query method, a more consistent range of improvements, despite having a slightly lower median.
\\
\\
Also, our statistical significance test results are shown in Table 2. The results indicate that both the Tree method and the Tree+Query method are statistically significant compared to the Origin method, with p-values of 0.027 and 0.022, respectively. The Cohen's d values for the Tree method and the Tree+Query method are 0.458 and 0.484, respectively, indicating a moderate effect size.

\begin{table}[ht]
    \centering
     \resizebox{\linewidth}{!}{
    \begin{tabular}{|l|c|c|c|c|}
    \hline
    \textbf{Comparison} & \textbf{Mean Difference} & \textbf{t-statistic} & \textbf{p-value (one-tailed)} & \textbf{Cohen's d} \\
    \hline
    Tree vs Origin (Total Score) & 0.416 & 2.05 & 0.027 & 0.458 \\
    Tree+QE vs Origin (Total Score) & 0.432 & 2.16 & 0.022 & 0.484 \\
    Tree vs Origin (Expert Score) & 0.700 & 1.73 & 0.050 & 0.387 \\
    Tree+QE vs Origin (Expert Score) & 0.700 & 2.40 & 0.013 & 0.538 \\
    \hline
    \end{tabular}
    }
    \caption{Statistical significance test results for each method compared to the baseline.}
    \label{tab:significance}
    \end{table}
\section{Limitations}
    Our study demonstrated the feasibility of integrating RAG with legal field, specifically in the context of Civil Law and Land Regulations.
    However, several limitations were observed that may guide future improvements and extensions.\\
    \\
    First, although CoT prompting facilitates more detailed and structured responses, both our experts in Civil Law and Land Regulations pointed out that while the overall direction of the responses were correct, there were some inaccuracies in the details. This issue likely stems from our design approach, which involved mimicking the reasoning patterns of legal experts by breaking down their thought processes into step-by-step CoT prompts. However, the more elaborate the response became, the greater the chance for subtle legal inaccuracies to emerge within specific reasoning steps. To address this issue, future work may consider introducing a mid-sized, legally trained model as a secondary calibration module. This legal expert model would take the generated responses as input and perform fine-grained corrections to ensure the accuracy of legal details.\\
    \\
    Second, while the system performed consistently and outperformed the baseline in Civil Law, the results in Land Regulations were less accurate. Our experts in Land Regulations hightlghted that the scenarios from Land Regulations tended to be more complex and often involved multiple legal provisions. According to their feedback, \enquote{Although the context of the question typically relates to Land Regulations, the actual legal disputes and reasoning often require knowledge that across multiple areas of law, rather than relying solely on the Land Act or a limited set of legal articles.}The detailed system responses and expert feedback for the Land Regulations section are included in the Appendix for reference. This suggests that effective legal reasoning in such scenarios demands broader legal integration, which remains a challenge for the current system. However, integrating multiple legal domains into the retrieval process may significantly increase the number of relevant documents retrieved, potentially reducing the efficiency and accuracy of the language model's response, or even exceeding the context window limitations of the model.\\
    \\
    Third, both experts indicated that the responses from the system lacked deeper jurisprudential reasoning. Jurisprudence is the fundamental spirit and theoretical foundation that forms a nation's entire legal system or specific branches of law. The complexity and constant evolution of social phenomena make it impossible for legislation to fully regulate every situation. In such cases, jurisprudence serves to supplement gaps in statutory and customary law, and is often regarded as a valid source of law. However, because jurisprudential reasoning is not explicitly codified in legal texts, it cannot be retrieved by the system—yet it remains an essential tool for legal professionals in interpreting and applying the law to real-world cases. To address this issue, future work may consider incorporating relevant judicial decisions into the model's training data, provided that sufficient computational resources are available. By exposing the model to a wide range of precedent cases, it would be better equipped to generate responses that reflect not only statutory reasoning but also underlying jurisprudential principles, leading to more comprehensive and legally sound judgments in context-specific scenarios.

\section{Future Work}

    We also explore technical and methodological improvements that could be made to the system, as well as broader applications beyond the legal domain.
\subsection{Technical and Methodological Improvements}

    In addition to the limitations and potential extensions we discussed above, we identified several technical and methodological improvements that could enhance the system's performance and usability.\\
    \\
    A LLM trained on legal corpora including academic commentary, precedent cases, legal opinions, and answers to the legal questions from experts could develop a more nuanced understanding of legal reasoning and jurisprudence, which would help to solve almost all the limitations of the current system we mentioned above.\\
    \\
    However, considering the constraints of hardware and deployment environments, training a large-scale LLM on a large-scale legal corpus remains a significant challenge. This is the main reason why we chose to focus on how to improve the retrieval parts in RAG framework in this study. At most, we could afford to train a small- to medium-sized legal language model on a limited corpus. While such a model may not be sufficient to serve as the primary generator of the system, it could instead be used as a secondary module to calibrate and refine the detailed responses produced by our main system.

\subsection{Broader Applications}
    Considering that not all organizations have the hardware capacity to deploy a LLM, therefore RAG, providing a cost-effective alternative solution for those company. The RAG framework strikes a balance between computational efficiency and performance, making it particularly suitable for institutions such as small- to medium-sized companies, government agencies, or educational organizations that require domain-specific AI support but lack the resources to train or host a dedicated LLM.\\
    \\
    In this study, our algorithm demonstrates statistically significant improvements over the baseline RAG framework in legal reasoning tasks. Beyond legal applications, we expect that our tree-structured RAG system can be extended to other domains that require specialized knowledge, such as medicine, tourism, and education. For instance, in the medical field, the system could provide accessible health-related information or support basic diagnostic inquiries; in education, it could assist students with individual question answering or enable real-time Q$\&$A when combined with a speech-to-text (STT) system; in tourism, the system could offer detailed travel information or personalized destination recommendations. These are all use cases in which AI support is valuable, but access to large-scale computing infrastructure is limited.

\bibliographystyle{unsrt}
\bibliography{paper_ver}

\begin{thebibliography}{10}

\bibitem{10.5555/3495724.3496517}
Patrick Lewis, Ethan Perez, Aleksandra Piktus, Fabio Petroni, Vladimir Karpukhin, Naman Goyal, Heinrich K\"{u}ttler, Mike Lewis, Wen-tau Yih, Tim Rockt\"{a}schel, Sebastian Riedel, and Douwe Kiela.
\newblock Retrieval-augmented generation for knowledge-intensive nlp tasks.
\newblock In {\em Proceedings of the 34th International Conference on Neural Information Processing Systems}, NIPS '20, Red Hook, NY, USA, 2020. Curran Associates Inc.

\bibitem{vaswani_attention_2017}
Ashish Vaswani, Noam Shazeer, Niki Parmar, Jakob Uszkoreit, Llion Jones, Aidan~N Gomez, Ł~ukasz Kaiser, and Illia Polosukhin.
\newblock Attention is {All} you {Need}.
\newblock In I.~Guyon, U.~Von Luxburg, S.~Bengio, H.~Wallach, R.~Fergus, S.~Vishwanathan, and R.~Garnett, editors, {\em Advances in {Neural} {Information} {Processing} {Systems}}, volume~30. Curran Associates, Inc., 2017.

\bibitem{10.1145/3703155}
Lei Huang, Weijiang Yu, Weitao Ma, Weihong Zhong, Zhangyin Feng, Haotian Wang, Qianglong Chen, Weihua Peng, Xiaocheng Feng, Bing Qin, and Ting Liu.
\newblock A survey on hallucination in large language models: Principles, taxonomy, challenges, and open questions.
\newblock {\em ACM Trans. Inf. Syst.}, 43(2), January 2025.

\bibitem{karpukhin-etal-2020-dense}
Vladimir Karpukhin, Barlas Oguz, Sewon Min, Patrick Lewis, Ledell Wu, Sergey Edunov, Danqi Chen, and Wen-tau Yih.
\newblock Dense passage retrieval for open-domain question answering.
\newblock In Bonnie Webber, Trevor Cohn, Yulan He, and Yang Liu, editors, {\em Proceedings of the 2020 Conference on Empirical Methods in Natural Language Processing (EMNLP)}, pages 6769--6781, Online, November 2020. Association for Computational Linguistics.

\bibitem{qin-etal-2024-large}
Zhen Qin, Rolf Jagerman, Kai Hui, Honglei Zhuang, Junru Wu, Le~Yan, Jiaming Shen, Tianqi Liu, Jialu Liu, Donald Metzler, Xuanhui Wang, and Michael Bendersky.
\newblock Large language models are effective text rankers with pairwise ranking prompting.
\newblock In Kevin Duh, Helena Gomez, and Steven Bethard, editors, {\em Findings of the Association for Computational Linguistics: NAACL 2024}, pages 1504--1518, Mexico City, Mexico, June 2024. Association for Computational Linguistics.

\bibitem{xiong2020multihop}
Wenhan Xiong, Xiang~Lorraine Li, Srini Iyer, Jingfei Du, Patrick Lewis, William~Yang Wang, Yashar Mehdad, Wen tau Yih, Sebastian Riedel, Douwe Kiela, and Barlas Oğuz.
\newblock Answering complex open-domain questions with multi-hop dense retrieval, 2020.
\newblock arXiv preprint arXiv:2009.12756.

\bibitem{sarthi2024raptor}
Parth Sarthi, Salman Abdullah, Aditi Tuli, Shubh Khanna, Anna Goldie, and Christopher~D. Manning.
\newblock Raptor: Recursive abstractive processing for tree-organized retrieval, 2024.
\newblock arXiv preprint arXiv:2401.18059.

\bibitem{edge2024graphrag}
Darren Edge, Ha~Trinh, Newman Cheng, Joshua Bradley, Alex Chao, Apurva Mody, Steven Truitt, Dasha Metropolitansky, Robert~Osazuwa Ness, and Jonathan Larson.
\newblock From local to global: A graph rag approach to query-focused summarization, 2024.
\newblock arXiv preprint arXiv:2404.16130.

\bibitem{10.5555/3600270.3602070}
Jason Wei, Xuezhi Wang, Dale Schuurmans, Maarten Bosma, Brian Ichter, Fei Xia, Ed~H. Chi, Quoc~V. Le, and Denny Zhou.
\newblock Chain-of-thought prompting elicits reasoning in large language models.
\newblock In {\em Proceedings of the 36th International Conference on Neural Information Processing Systems}, NIPS '22, Red Hook, NY, USA, 2022. Curran Associates Inc.

\bibitem{wang2024e5}
Liang Wang, Nan Yang, Xiaolong Huang, Linjun Yang, Rangan Majumder, and Furu Wei.
\newblock Multilingual e5 text embeddings: A technical report, 2024.
\newblock arXiv preprint arXiv:2402.05672.

\bibitem{openai2025gpt4o_mini}
OpenAI.
\newblock Gpt-4o-mini models, 2025.
\newblock Available at: \url{https://platform.openai.com/docs/models/gpt-4o-mini} (accessed: 2025-04-12).

\end{thebibliography}

\appendix
\section{Appendix}

\subsection{Prompt Design}

\subsubsection{Task-Oriented Prompt}

\begin{table}[H]
\centering
\caption{任務導向提示（Task-Oriented Prompt）}
\begin{tabular}{|p{14cm}|}
\hline
\textbf{Task-Oriented Prompt:} \\[0.5em]
你將獲得以下兩個資訊：\\
- \textbf{法律問題:} \{query\} \\
- \textbf{檢索內容:} \{context\} \\[0.5em]

你是一個專業的法律顧問機器人，需根據法律問題和檢索內容提供準確且清晰的回答。\\[0.5em]

\textbf{回答步驟：} \\
1. 確認關聯性：若問題與檢索內容相關，則分析並回答；若無關，則忽略。\\
2. 使用 Thought → Action → Observation 框架：\\
\quad Thought：確認問題可否根據檢索內容回答。\\
\quad Action：從檢索內容中提取所有你認為解答所需之關鍵檢索內容或細化問題以確保準確性。\\
\quad Observation：提供清晰、準確的回答。\\[0.5em]

\textbf{回應規則：} \\
- 僅使用檢索內容，不使用內部知識。\\
- 如資訊不足，應明確指出缺少部分。\\
- 無法回答時，直接回應：「我沒有足夠的相關資訊來回答這個問題」。 \\
\\
\hline
\end{tabular}
\end{table}

\subsubsection{CoT Prompt}
\begin{table}[H]
\centering
\caption{Chain of Thought 推理提示（CoT Prompt）}
\begin{tabular}{|p{14cm}|}
\hline
你將獲得以下資訊：\\
- 原始問題：\{query\} \\
- 檢索內容：\{context\} \\[0.5em]

你是一位專業法律顧問機器人，請依照以下【Chain of Thought（CoT）推理步驟】完整邏輯分析並作答：\\[0.5em]

\textbf{【Step 1】明確界定核心法律問題} \\
- 依原始問題清楚界定本案需解決的核心法律問題。\\[0.5em]

\textbf{【Step 2】概念與法律地位辨析（此步驟必須執行）} \\
- 針對檢索內容中出現的\textbf{專有法律概念、主體身分或專業術語}進行清楚辨析與定義。\\
- 如發現有\textbf{名稱相近但法律效果或法律地位不同的概念}（例如：無行為能力人 vs 限制行為能力人），請完整區分並標示，避免混淆。\\
- 所有專業概念請逐一定義，並說明彼此區別及法律效果差異。\\[0.5em]

\textbf{【Step 3】提取關鍵法律事實與條文} \\
- 條列與核心法律問題相關的法律事實、法條或案例。\\[0.5em]

\textbf{【Step 4】逐步推理與法律適用} \\
- 依照抽取出的事實與條文，進行條理清晰的法律推理。\\
- 如有爭議或多種見解，請分別說明。\\[0.5em]

\textbf{【Step 5】得出法律結論} \\
- 明確回答本案的法律問題。\\
- 若檢索內容不足，請直接回答：「依目前檢索內容，無法完整回答。」\\[0.5em]

\textbf{⚠️ 特別規則：} \\
- 嚴格依照檢索內容推理，禁止引入外部知識或假設。\\
- 法律條文請完整標示「XX法第 X 條第 X 項」。\\
- 保持用語精確，嚴防法律概念混淆。\\
\\
\hline
\end{tabular}
\end{table}

\subsubsection{Query Extraction Prompt}
\begin{table}[H]
\centering
\caption{法律助理提示（Legal Assistant Prompt）}
\begin{tabular}{|p{14cm}|}
\hline
你是一位專業的法律助理，專責協助律師從案件事實或法律問題中，提取出核心法律事實、主要法律爭點，以及涉及的法律條文或當事人主張，並依下列結構化格式整理：\\[0.5em]

\textbf{【背景核心事實】} \\
- 條列案件或題目中出現的客觀事實，包括當事人身分、行為、處分或其他法律上重要事實。\\[0.5em]

\textbf{【主要法律爭點】} \\
- 條列案件或題目中需要解決的核心法律問題，包括法律關係、權利義務歸屬、法律行為效力、構成要件等。\\[0.5em]

\textbf{【概念與專有名詞辨析】（如無混淆疑慮可省略）} \\
- 若案件中出現易混淆或法律效果不同之相似名詞（如「無行為能力人」與「限制行為能力人」），請完整列出並分別定義，明確說明彼此差異與法律效果。\\[0.5em]

\textbf{⚠️ 以下兩個區塊僅在有內容時產出，若無內容請完全省略，不得填「無」：} \\[0.5em]

\textbf{【當事人或機關主張】} \\
- 條列當事人或機關在案件中明確表達的主張、法律見解或立場。\\
- 特別注意：若案件有引用法律條文，請完整標示為：\\
\textbf{「XX法第X條第X項」}，不得簡化、不得省略法條名稱與條次。\\[0.5em]

\textbf{【涉及的法律條文】} \\
- 若案件事實或法律問題內容中直接提及的重要法律依據，完整標示為：\\
\textbf{「XX法第X條第X項」}。\\[0.5em]

\textbf{⚠️ 特別注意：} \\
- 嚴禁加入任何推論、補充、法條或自行延伸，僅可案件提取描述內容。\\
- 請保持結構清晰、條列明確，專業且客觀，符合法律書面標準。\\
- 所有法律條文及專有名詞須確保完整、正確，避免概念混淆。\\[0.5em]

以下為案件事實或法律問題內容：\{query\} \\
\\
\hline
\end{tabular}
\end{table}

\subsection{Example QA Pairs}

\subsubsection*{Q1. 甲授與代理權予乙，欲乙代理甲向丙購買A物一件，若乙係以自己名義向丙購買A物一件，此買賣效力是否及於甲？又丁授與代理權予戊，欲戊代理丁向庚購買B物一件，並給予戊授權書。在戊未購買前，丁發現戊不能信賴，遂向戊表示撤回前開授權，惟忘記將授權書索回，旋戊出示授權書，以丁之名義向庚購買B物一件，則該買賣效力是否及於丁？}

\textbf{Generated Answer:} \\
Final Result:

【Step 1】明確界定核心法律問題
- 本案需解決的核心法律問題為：甲與丁在授權代理的情況下，乙與戊以各自名義購入物品的交易有效性及其對於甲和丁的效力。

【Step 2】概念與法律地位辨析
- **代理權**：指一方（授權者）授予另一方（代理人）代為處理事務的權利。
- **無行為能力人**：依據法律，無法獨立從事法律行為的人，其一切法律行為均無效。
- **限制行為能力人**：依據法律，雖然可以從事法律行為，但在某些行為上需要他人的同意或許可（如法定代理人的允許）。
- **善意第三人**：對於法律行為抱持善意並無過失的人，法律特別保護其權益，不受他人不法行為影響。

【Step 3】提取關鍵法律事實與條文
- 甲授與代理權於乙，乙向丙購買A物，若乙以自己名義購買，根據《民法》第104條，代理人所為之法律行為，其效力應由授權者承擔，假如代理人沒有超越授權的情況。
- 丁授予戊代理權並賦予授權書，但若丁發現戊不可信賴並撤回授權，根據《民法》第107條，代理權之限制及撤回不得對抗善意第三人，且《民法》第109條規定代理人不應留置授權書，需交還授權者。

【Step 4】逐步推理與法律適用
- **針對甲和乙的情形**：
  - 乙若在其授權範圍內以甲的名義購買A物，該買賣效力及於甲。
  - 若乙以自己名義購買，則該買賣一般視為不符合代理的要求，如無法證明乙在行為時的權限，則不會對甲產生法律效力。

- **針對丁和戊的情形**：
  - 丁在知道戊不可靠的情況下撤回授權，但未索回授權書，根據《民法》第107條，戊依舊可以以丁的名義與庚交易，因為戊在第三方（庚）的善意情況下使交易有效。
  - 若庚是善意的，他無法知道戊的授權已被撤回，則此交易對丁依然具有約束力。

【Step 5】得出法律結論
- 乙以自己名義購買A物的交易不及於甲，因未在授權範圍內。而戊向庚購買B物的交易，因戊未撤回的授權書對於善意的第三人庚有效，故該交易效力及於丁。

\bigskip

\subsubsection*{Q2. 甲因患有阿茲海默症，時常認不得人胡言亂語，經家人向法院聲請，而受監護宣告。某日，甲神智清醒能正常表達，便自行到住家附近，由乙獨資經營的餐廳點餐吃飯，用餐完畢後，甲不想付錢，遂主張自己受監護宣告，契約無效，故可以不用付錢，隨即離開。
1-1甲拒絕支付餐費的主張是否有理由？}

\textbf{Generated Answer:} \\
Final Result:

【Step 1】明確界定核心法律問題
- 本案的核心法律問題是：甲是否可以主張契約無效以拒絕支付餐費，因其受監護宣告的身份。

【Step 2】概念與法律地位辨析
1. **無行為能力人**：根據《民法》第15條，受監護宣告之人視為無行為能力人，因此其所為的意思表示均為無效。
2. **限制行為能力人**：限制行為能力人雖然可以進行某些法律行為，但必須經過法定代理人或輔助人的同意，否則該行為的效力不完全。
3. **監護宣告**：根據《民法》第14條，法院可以依據聲請對精神障礙或其他心智缺陷的個體（如甲）進行監護宣告，這表示該人於法律行為上不具有能力。
4. **輔助宣告**：針對精神障礙或心智缺陷輕微的情況，法院也可進行輔助宣告，該人仍具備一定法律行為的能力，但需經輔助人的同意。

法律效果方面，無行為能力人所成立的契約無效，而限制行為能力人未經同意所成立的契約需經法定代理人或輔助人承認才能生效。

【Step 3】提取關鍵法律事實與條文
1. 甲因阿茲海默症，已經受到監護宣告（根據《民法》第15條）。
2. 甲在神智清醒時，獨自前往餐廳消費。
3. 甲用餐後拒絕支付，主張契約無效。

相關條文：
- 《民法》第15條：受監護宣告之人，無行為能力。
- 《民法》第81條：限制行為能力人承認所訂立之契約具有同一效力。
- 《民法》第82條：限制行為能力人所訂立之契約，未經承認前，相對人得撤回之。

【Step 4】逐步推理與法律適用
首先，甲已受到監護宣告，根據《民法》第15條，其法律行為已被視為無效。因此，甲所主張的契約無效，是有法理基礎的。

其次，即使在其神智清醒時所表現出來的行為（如點餐用餐）似乎具備行為能力，但根據《民法》第15條規定，受監護宣告者即使在清醒時，其法律行為仍可視為無效。

再者，根據《民法》第81條與第82條，限制行為能力人的契約需經法定代理人之承認方能生效，而甲的狀態不同於限制行為能力人，他直接被認定為無行為能力人，因此不需要經任何承認就可以主張契約無效。

然而，若甲在神智清醒期間進行合同或語言上發出意思表示，一般不是基於缺乏理解的情況下所做出，而此舉本身可能引發與餐廳的合法交易關係的社會期待及法律期待。因此，儘管甲以受監護宣告為由主張契約無效，仍需考慮其行為的社會倫理與實務操作。

【Step 5】得出法律結論
根據上述推理，甲以「受監護宣告」主張契約無效的理由是成立的，即甲可以拒絕支付餐費。該契約依《民法》第15條規定視為無效，故甲的拒絕支付是有法律依據的。

\bigskip

\subsubsection*{Q2. 甲因患有阿茲海默症，時常認不得人胡言亂語，經家人向法院聲請，而受監護宣告。某日，甲神智清醒能正常表達，便自行到住家附近，由乙獨資經營的餐廳點餐吃飯，用餐完畢後，甲不想付錢，遂主張自己受監護宣告，契約無效，故可以不用付錢，隨即離開。
1-2乙對於餐費是否有民法上的權利可以向甲主張？}

\textbf{Generated Answer:} \\
Final Result:

【Step 1】明確界定核心法律問題
本案的核心法律問題在於：甲因受監護宣告而是否具有支付餐費的法律責任，乙是否可以對甲主張民法上的權利。

【Step 2】概念與法律地位辨析
1. **受監護宣告**：根據《民法》第15條，受監護宣告之人為無行為能力。即該類人無法獨立從事法律行為。
2. **無行為能力人**：該類人由于精神障礙或其他心智缺陷，不能進行意思表示或辨識意思表示的效果，因此法律行為不生效力。
3. **限制行為能力人**：此類人具有一定的行為能力，但某些行為仍需經法定代理人的同意。
4. **法定代理人**：指根據法律或法院的指定，能代表無行為能力人或限制行為能力人進行法律行為的個體。

這些概念相互區分之處在於，無行為能力人（如受監護宣告者）其所訂立之契約全然無效，而限制行為能力人則在一定條件下仍可訂立有效的法律行為。

【Step 3】提取關鍵法律事實與條文
1. 甲因患有阿茲海默症而經法院聲請受監護宣告，根據《民法》第15條第1項，甲為無行為能力人。
2. 甲在某日神智清醒能正常表達，獨自到餐廳就餐並點餐，飲食後拒絕支付餐費，主張契約無效。
3. 根據《民法》第81條，限制行為能力人於限制原因消滅後，承認其所訂立之契約者，承認與法定代理人承認相同。

【Step 4】逐步推理與法律適用
甲的行為本質上是違反了支付餐費的義務。儘管有時甲的精神狀態較為清醒，但根據《民法》第15條，因受監護宣告而使甲視為無行為能力，因此其所訂立的就餐契約不具法律效力。乙雖然在甲表示用餐並確認之後提供服務，但因甲的無行為能力，故乙無法對甲主張支付的權利。

根據《民法》第79條規定，限制行為能力人若未得法定代理人之允許，其所訂立的契約未經承認前他方可撤回。因此，乙無法就甲的行為要求支付餐費。

此外，根據《民法》第82條，限制行為能力人所訂立之契約，未經承認前，相對人可撤回。可見，甲既然被視為無行為能力人，乙的要求是無效的。

【Step 5】得出法律結論
依據上述推理，乙對於甲的餐費主張無法成立。因甲因受監護宣告為無行為能力，故其與乙間的契約不具有法律效力，因此無須支付餐費。

\bigskip

\subsubsection*{土地法第12條規定：「(第一項)私有土地，因天然變遷成為湖澤或可通運之水道時，其所有權視為消滅。（第二項）前項土地，回復原狀時，經原所有權人證明為其原有者，仍回復其所有權。」請闡釋本條第二項規定內容。}

\textbf{Generated Answer:} \\
Final Result:

【Step 1】明確界定核心法律問題
- 本案需解決的核心法律問題是：土地法第12條第2項所規定的「回復原狀時，經原所有權人證明為其原有者，仍回復其所有權」的具體含義及法律效力。

【Step 2】概念與法律地位辨析
- **私有土地**：指屬於個人或法人所有的土地，與公有土地相對。私有土地的所有者擁有完全的使用、收益、處分權和排除他人干涉的能力（土地法第765條）。
  
- **所有權**：指對物的法律支配權，所有權人可以自由使用、收益和處分其所擁有的財產。若法律或合同未另行規定，其權利為絕對的。

- **回復原狀**：此概念在法律上通常指物的狀態恢復至自然或法定的初始狀態。在土地法第12條情境中，若土地因天然變遷變成湖澤或水道，當這片土地「回復原狀」時，意味著水位下降或自然條件改變，使土地重現為可用的陸地。

- **證明為其原有者**：意味著原土地所有權人需要提供足夠的證明，證明其確為該土地的合法所有者。在法律上，證明責任通常在聲明擁有權一方。

【Step 3】提取關鍵法律事實與條文
- 條文：土地法第12條第2項規定：「前項土地，回復原狀時，經原所有權人證明為其原有者，仍回復其所有權。」
- 關鍵事實：私有土地因天然變遷變成湖澤或可通運之水道後，所有權視為消滅；當該土地回復為陸地，原所有權人有權證明其所有權並重新獲得所有權。

【Step 4】逐步推理與法律適用
- 根據土地法第12條第1項的規定，私有土地因天然變遷成為湖澤或水道時，其所有權視為消滅，這表明在特定天然災害或變遷下，土地的所有權並非永久消失，而是暫時因自然因素被撤銷。

- 根據第2項的規定，在該土地回復為可使用的陸地時，只要原所有權人能提供證明，證明自己是該土地的合法擁有者，就可以重新獲得對該土地的所有權。這一條款保護了原所有權人的權益，確保在法律上他們依然有可能恢復失去的財產權。

- 因此，如果某個自然或環境的變化使土地轉變為湖澤或水道，雖然所有權在法理上被視為消滅，當土地恢復其原狀，且原所有權人做好證明時，即可回復其所有權，這展示了法律對於所有權的保護及其在特殊情況下的響應。

【Step 5】得出法律結論
- 根據土地法第12條第2項的規定，原所有權人在土地因天然變遷而喪失所有權後，若該土地後來再次回復為可使用的狀態，並且能夠證明其原始所有者身份，即可以回復其所有權。這一制度設計有助於保障原所有權人的合法權益。

\end{document}